%Paper: nucl-th/9407029
%From: Helmut.Hofmann@Physik.TU-Muenchen.DE
%Date: Tue, 19 Jul 94 12:34:21 +0200

\input hivcom
\input hof_lit
\input hof_book
\input aheq.tex
\input ahdef.tex

%%%%%%%%%%%

\centerline{{\headl On the macroscopic limit of nuclear dissipation}
\footnote{$^1$}{Supported in part by the Deutsche Forschungsgemeinschaft
under 436 UKR 113-15}
}
\bigskip %\bigskip
\centerline{by}
\bigskip %\bigskip
\centerline{{\bf Valery I. Abrosimov}
\footnote{$^2$}{Permanent address: Institute for Nuclear Research, Kiev,
Ukraine}
{\bf and Helmut Hofmann}}
\medskip
\centerline{Physik-Department, T30, TUM, D-85747 Garching}

\vskip 1.5cm
%%%%  ABSTRACT
\parindent=20pt

\par{\narrower\narrower\noindent
The Landau-Vlasov equation is applied to a slab of width $L$. This
geometry is introduced to simulate somehow the finiteness of real
nuclei but to allow for analytical solutions, nevertheless. We focus on
the damping of low frequency surface modes and discuss their friction
coefficient. For this quantity we study the macroscopic limit as
defined by $L\to \infty$. We demonstrate that the same result can be
obtained for finite $L$ by applying an appropriate frequency smoothing,
if only the smearing interval is sufficiently large. The apparent, but
important consequences are discussed which this result will have for the
understanding of the nature of dissipation in real nuclei.
\par}
\parindent=20pt
\vskip 1.0cm
%\centerline{\date}
\centerline{22.4.1994}
\vskip 0.5cm
\centerline{to appear in Nuclear Physics}
\topskip=10pt
%%%%%%%%%%%%%%%%% TEXT %%%%%%%%%%%%%%

\vskip 1.0cm
{\bf \num Introduction}
\bigskip
\ref{hirorev}
\edef\nhirorev{\the\refnumber}
\ref{higoroobn}
\edef\nhigoroobn{\the\refnumber}

The nature of nuclear dissipation is still not completely clarified. For
slow collective motion, the case on which we shall concentrate here,
the only experimental information we seem to have is from fission data.
Detailed analyses appear to favour (see [\nhirorev] and [\nhigoroobn])
the results of one-body friction over the ones
which relate to two body viscosity. On the other hand there can be no
doubt that for the finite temperatures or excitations, for which the
measurements are done, two body collisions cannot be neglected.

\ref{ranswiat}
\edef\nranswiat{\the\refnumber}
\ref{abran}
\edef\nabran{\the\refnumber}
\ref{abkol}
\edef\nabkol{\the\refnumber}
\ref{koonrand}
\edef\nkoonrand{\the\refnumber}
This feature is quite puzzling as commonly the wall formula
[\nranswiat] is derived
{\it disregarding} any traces of a collision term. However, all the
derivations we know of involve the {\it macroscopic limit}: Often one
refers to the case of a semi-infinite system from the start, like it is
done in applications of the Landau-Vlasov equation [\nabran, \nabkol],
%((HASSE ??))
or one first looks at a finite system and lets its size go to infinity
afterward, as is done for instance in [\nkoonrand].

\ref{reif}
\edef\nreif{\the\refnumber}

The question we are touching here is very closely related to how one
introduces irreversibility into a dynamical description. For
an infinite system the level spacing of micro-states is arbitrarily
small rendering any recurrence time much larger than the times
relevant for the change of observable quantities. For a finite nucleus,
the situation may be very different. There the spectra of
one-particle one-hole excitations are {\it not dense}, but these are
exactly the ones which commonly one adverts to when one
describes collective motion. Indeed, the effects of collisions are
suppressed for scattering of nucleons at the Fermi energy, simply
because of the Pauli principle. For these reasons the nucleus was
sometimes compared to a "Knudsen gas" (for a definition of the latter
see e.g.[\nreif]) when its macroscopic limit was to be interpreted physically.

We claim that this picture is inadequate and we want to prove our
conjecture by looking at an analytically solvable model of a slab. We
shall be able to perform the transformation to the macroscopic limit in
two ways, firstly by simply letting the size of our system go to
infinity, and secondly by performing energy or frequency smoothings
while keeping the size of the system finite.  As we shall see, in the latter
case the smoothing interval must be chosen quite large. It is here
where for the case of a finite system statistical concepts are
introduced, and thus irreversibility.  To which extent the
necessary conditions are actually fulfilled in a real situation will be
discussed in our last section.

The dynamics inside the bulk will be described with the Landau-Vlasov
equation. It will have to be accomplished by an appropriate treatment
of the surface. As outlined in sect.2 we will use the
framework of linear response theory. Following common schemes we then
will identify in sect.3 the friction coefficient for surface modes. The
central issue of our paper will be discussed in sect.4. There it will
be proven that the macroscopic limit can be simulated or obtained by
frequency smoothing as well.  Relations to the
common procedures of applying energy smoothings to finite systems will
be discussed both in sect.4 well as in sect.5.

\vskip 10pt
\goodbreak
{\bf \num Surface response function for a slab model}
\bigskip
\ref{feibelman}
\edef\nfeibelman{\the\refnumber}
\ref{ivan}
\edef\nivan{\the\refnumber}

In [\nabkol] a semiclassical approach has been proposed to discuss
surface modes for a semi-infinite system. We extend this model
to slab geometry in the hope that in this way we may simulate
better some effects of the nucleus being a {\it finite system}. Our
system will consist of a Fermi liquid bound by the two surfaces
\equorder{surfeq}
\surfeq
\edef\nsurfeq{\the\equnumbera}
parallel to the $xy-$ plane and, in equilibrium, a distance $L$ apart.
Here and below the subscript $\perp$ denotes a
vector perpendicular to the z-axis, i.e. ${\vec r}_\perp=(x,y,0)$.
Notice please, that this choice of the boundary means that both
surfaces move {\it in phase}. Such a situation has been studied before
in [\nfeibelman] within a (quantal) time-dependent self-consistent
field approximation and in [\nivan] applying  Landau-Vlasov dynamics.
In both cases it had been argued that the "in phase" motion for the slab
corresponds to the dynamics of low-lying collective vibrations of
finite nuclei---in contrast to breathing modes which one would like to
associate to an "out of phase" motion of the slab.
\ref{sije}
\edef\nsije{\the\refnumber}
\ref{kidhofiva}
\edef\nkidhofiva{\the\refnumber}

The quantity $Z({\vec r}_\perp,t)$ describes the local displacement of
the surfaces from its positions in equilibrium. In the sequel we want
to call $Z({\vec r}_\perp,t)$ our macroscopic variables. Its definition
is in accord with standard procedures for collective motion (see e.g.
[\nsije] and [\nkidhofiva]); we shall return to this question below
when we come to discuss response functions or the energy transferred to
the system by an external force.

A change in $Z$ will induce motion of the fluid particles inside the
slab. The latter shall be represented by the distribution function
$\dnrpt$ in the particle's  phase space.  We will assume that
deviations from the equilibrium distribution  $n_0$ are small such that
they can be described by a linearized Vlasov equation. At zero
temperature this distribution then simply is a function of the single
particle energy $n_0=n_0(\epsilon(\vec r,\vec p))$
\equorder{vlasoveq}
\vlasoveq
\edef\nvlasoveq{\the\equnumbera}
The ${\cal F}_0$ stands for ${\cal F}_0 = F_0 h^3  / (4 \pi m p_F)$,
where $F_0$ is the Landau parameter.

\ref{bekha}
\edef\nbekha{\the\refnumber}

This equation, valid inside the slab at $-L/2 < z <L/2$, sometimes
referred to as the "bulk" of the system, has to be accomplished by
appropriate boundary conditions at the moving surfaces (\nsurfeq). Following
[\nbekha] and [\nivan]  these conditions can be written as:
\equorder{mirrefl}
\mirrefl
\edef\nmirrefl{\the\equnumbera}
if we assume that the particles bounce back from the walls
under the requirement of exact "mirror reflection". The collision with
the wall is considered to happen  elastically in the frame of the
moving surface. For details see [\nbekha], in particular eq.(13).
These two equations would suffice to fully specify the dynamics of the
system, if the $Z({\vec r}_\perp,t)$ would be considered a truly {\it
external} parameter.
However, we want to look at a different situation.
Like in the case of a finite nucleus, the motion of the surface must be
determined "self-consistently"---in the sense that it has to come from
an exact balance of the {\it internal} forces acting on a {\it free} surface.
Following [\nivan, \nabkol] this can be achieved by asking for the following
"subsidiary condition":
\equorder{balforc}
\balforc
\edef\nbalforc{\the\equnumbera}
\ref{lalifhyd}
\edef\nlalifhyd{\the\refnumber}
\ref{lalifphyskin}
\edef\nlalifphyskin{\the\refnumber}
Here, $\sigma$ is the surface tension and $P_{zz}(\vec r,t)$ is the
normal component of the momentum flux tensor, see [\nlalifphyskin].
Actually, in (\nbalforc) we have included an {\it external} force,
namely $F_{\rm ext}({\vec r}_\perp,t)$. This has been done simply to
enable us identify later the response of the system to such an external
source and thus to benefit from the tools of linear response theory. If
the system were to move freely, i.e. with a $F_{\rm ext}({\vec
r}_\perp,t) = 0$, eq.(\nbalforc) would become identical to the form
discussed in chap. 7 of [\nlalifhyd].  For obvious reasons, the $F_{\rm
ext}({\vec r}_\perp,t)$ shall be written in the following form
\equorder{extforc}
\extforc
\edef\nextforc{\the\equnumbera}
with $\om$ being the frequency of the external perturbation and
$\eps = +0$ representing an infinitesimally small quantity
to guaranty that the external field  is turned on adiabatically at
$t=-\infty$. Please notice, that the boundary condition (\nbalforc) are
chosen such that the external force excites a motion of both surfaces
in phase.

Of interest are those solutions of eqs.  (\nvlasoveq)-(\nmirrefl) which
are tailored to the special form of the external force (\nextforc).
First of all we may anticipate that in its dependence on ${\vec
r}_\perp,t$  the $Z$ follows the one of the external perturbation.
Thus we may write:
\equorder{macvar}
\macvar
\edef\nmacvar{\the\equnumbera}
For the $\dnrpt$ we seek for a Fourier expansion of the form:
\equorder{disfunc}
\disfunc
\edef\ndisfunc{\the\equnumbera}
with the $f_{k_\perp,\omega}(z,\vec p)$ expressed by:
\equorder{fourexp}
\fourexp
\edef\nfourexp{\the\equnumbera}

\ref{lan}
\edef\nlan{\the\refnumber}

\ref{fom}
\edef\nfom{\the\refnumber}

\ref{iv}
\edef\niv{\the\refnumber}

In this way the problem can be reduced to a solution of an algebraic
equation for the components $f_{k_\perp,\omega}(k_n,\vec p)$.
The method employed here is similar to the one
proposed first in [\nlan] and used later on in refs.[\nfom,\niv,\nabran].
Therefore we do not want to present any further details.
After some straightforward calculations the Fourier components of
(\nfourexp) are easily found to be:
\equorder{compfourexp}
\compfourexp
\edef\ncompfourexp{\the\equnumbera}
where
\equorder{kn}
\kn
\edef\nkn{\the\equnumbera}

In writing the solution like (\ncompfourexp) we have put the Landau
force equal to zero $(F_0=0)$, meaning to treat the dynamics of the
bulk as the one of free particles. We are going to make this
approximation also in the sequel where we shall focus on the study of
the friction coefficient. This approximation is not only done for the
sake of simplicity. Indeed, originally the wall formula had been derived just
for such a system with the particles being treated as a Fermi gas
[\nranswiat]. Moreover, it was found in [\nabkol] for the case of a
semi-infinite Fermi liquid that the interaction of particles has only
negligible effects on the dissipative properties of surface
excitations. It seems plausible then to make the same hypothesis for
our slab model although further studies might be appropriate.

\ref{holet}
\edef\nholet{\the\refnumber}
\ref{hofiva}
\edef\nhofiva{\the\refnumber}
Using (\ncompfourexp) we can calculate the normal component of the
momentum flux tensor on the moving free surfaces of the system, see the
l.h.s. of (\nbalforc). To linear order in the surface
displacement $\zrt$  we find:
\equorder{nocomom}
\nocomom
\edef\nnocomom{\the\equnumbera}
with
\equorder{nocomomre}
\nocomomre
\edef\nnocomomre{\the\equnumbera}
and
\equorder{normcomp}
\normcomp
\edef\nnormcomp{\the\equnumbera}
(Notice that we neglect the spin degeneracy factor).
The physical interpretation of this function is simple. It
parametrizes how the system of the fluid particles {\it responds}
to a change in the surface. In this sense its meaning  is analogous
to the one of the "intrinsic response function" used often for
collective motion of finite nuclei (see e.g. [\nholet, \nsije,
\nkidhofiva]). Also the structure of this function resembles somehow
the one found for finite systems (see. e.g. eq.(7)
of [\nhofiva]). This is particularly striking for $k_\bot = 0$, for which
the $k_nv_z$ in the energy denominator takes the role of a "discrete
frequency". It is this feature which below will allow us to apply
smoothing procedures to obtain the macroscopic limit. As we shall see,
a finite $k_\bot$ does not not necessarily destroy this feature.

We may now parametrize the solutions (\nmacvar) in terms of functions
which measure the response to the external field $F_{\rm ext}$.
Following standard notation the latter may be defined as ([\nsije] and
[\nkidhofiva]):
\equorder{defresp}
\defresp
\edef\ndefresp{\the\equnumbera}
For the present purpose the following form is more convenient, in
particular if we want to be in accord with the notation used before in
papers on semi-infinite systems [\nabkol]:
\equorder{surfrespchi}
\surfrespchi
\edef\nsurfrespchi{\the\equnumbera}
The $\chipr$ and $\chidpr$ are the real and imaginary parts
of the response function $\chinp$, namely $\chinp= \chipr + i \chidpr$,
for real $\om$. With the help of (\nbalforc) and (\ncompfourexp) the
$\chinp$ is found to be:
\equorder{surfrespfchi}
\surfrespfchi
\edef\nsurfrespfchi{\the\equnumbera}
The poles of this function determine the possible modes of our
system which get  excited by the external field introduced in
eq.(\nbalforc).

\vskip 10pt
\goodbreak
{\bf \num Friction coefficient for collective motion}
\bigskip
Friction is a measure for energy loss, intimately interwoven both
with the concept of {\it irreversibility} as well as with {\it statistical
excitations}. Before we come to discuss this latter aspect in
the next section, let us identify here those parts of the effective
forces which parametrize energy loss in more general context.

The rate of work done by the external force on the system is
proportional to this force, here denoted by $F_{\rm ext}$, times the
velocity of the quantity which parametrizes the excitation of the
system, and which in the present case must be proportional to $Z$.
The rate of transfer of energy from the external force (\nextforc) to the
system, measured per unit element of the surface, is determined by:
\equorder{enertrd}
\enertrd
\edef\nenertrd{\the\equnumbera}
Here a {\it negative} sign means transfer of energy {\it to the
system},
in accord with standard notations in the literature. For the nuclear case see
e.g. chapter 8 of [\nsije] or eq.(57)  of [\nkidhofiva].

\ref{abkolnp}
\edef\nabkolnp{\the\refnumber}

For the following it is very useful to express $dE /dt$
entirely by the various parts of the response function and the collective
variable $Z$. Indeed, with the help of (\nsurfrespchi)
the external field $F_{\rm ext}$ can be removed, to give finally:
\equorder{enertrchi}
\enertrchi
\edef\nenertrchi{\the\equnumbera}
This expression will allow us to identify uniquely the dissipative
force, if we only follow closely the arguments of put forward in
[\nkidhofiva]. It is important to realize that the expression
(\nenertrchi) is valid both with as well as without an external force.
For $F_{\rm ext}({\vec r}_\perp,t) \neq 0$ the system can be forced to perform
strict harmonic oscillations. If we then average over one period the
first term on the right of (\nenertrchi) vanishes. The second one will
survive, however, indicating clearly that it is this term which is to
be associated with {\it irreversible} transfer of energy. This
feature will prevail even if the external field is put zero,
in which case the (free) motion of the surface will be {\it damped}
(and thus will deviate from a pure oscillatory behaviour).

The equation of motion for the free surface could be obtained from
energy conservation (see [\nholet], [\nkidhofiva] and
[\nabkolnp]), that is to say by putting the left hand side of
eq.(\nenertrchi) equal to zero. This shall not be done here. We may refer
to [\niv], however, where the corresponding dispersion relation has
been studied.  We like to concentrate on the coefficient of dissipation
and will calculate the latter in the so called "zero
frequency limit" [\nkidhofiva]. This is a meaningful approximation
whenever the frequency of collective motion turns out to be
sufficiently small. In [\nabran,\nabkol] it was found
that this limit does apply to the case of the semi-infinite
Fermi liquid, and we want to take over this hypothesis
to the model we are studying here. We may note, indeed, that this
assumption is in accord with the dispersion equation found in [\niv].

In (\nenertrchi) the quantity of interest then is the coefficient in
front of the $(\dot Z)^2$, which we want to call $\gamkomv$ in the sequel:
\equorder{enertrfacchi}
\enertrfacchi
\edef\nenertrfacchi{\the\equnumbera}
Using (\nsurfrespfchi) it can be expressed as
\equorder{enertrfac}
\enertrfac
\edef\nenertrfac{\the\equnumbera}
Inspecting the expression (\nnormcomp) for the intrinsic response
function
$\chi_{int}$, one realizes that the integration over the momentum space
can be done analytically [\nabkol], if for $\dnode$ one uses the simple
expression:
\equorder{delfunc}
\delfunc
\edef\ndelfunc{\the\equnumbera}

In this way for $\gamkomv$ the following formula is obtained:
\equorder{enertrfacp}
\enertrfacp
\edef\nenertrfacp{\the\equnumbera}
We have introduced the quantity
\equorder{fricwf}
\fricwf
\edef\nfricwf{\the\equnumbera}
which is the wall formula of nuclear friction [\nranswiat] (neglecting
spin degeneracy), and we have used the following short hand notations
\equorder{dcwv}
\dcwv
\edef\ndcwv{\the\equnumbera}
\equorder{dsnwv}
\dsnwv
\edef\ndsnwv{\the\equnumbera}
and
\equorder{funcwsn}
\funcwsn
\edef\nfuncwsn{\the\equnumbera}

\ref{kolsie}
\edef\nkolsie{\the\refnumber}
Within the zero frequency limit mentioned earlier friction
is obtained by evaluating $\gamkomv$ at $\omega = 0$:
\equorder{fric}
\fric
\edef\nfric{\the\equnumbera}
\ref{svvm}
\edef\nsvvm{\the\refnumber}
It is reassuring for our interpretation of the function $\ptilkom$ that
in this way we are in accord with a common definition of diispation
[\nholet, \nkoonrand, \nkolsie]. The value of the friction
coefficient can now be calculated from (\nenertrfacp) to give:
\equorder{fricp}
\fricp
\edef\nfricp{\the\equnumbera}
It can be said that the same result can be read off from the dispersion
relation found in [\niv] (see there eq.(59)). For $k_\perp =0$ for
which the friction coefficient (\nfricp) vanishes, it is also in accord
with the excitation function found in [\nsvvm], see therein
eqs. (25),(39) and (40).

At this stage it is worth while to remind the reader of one of our
basic assumptions. We have looked at solutions of the fundamental
equations of motion whose time dependence in a sense was determined by
our ansatz for the perturbation (\nextforc). Remember the
small imaginary part $\eps$ introduced there. In general this $\eps$
will then be much smaller than the spacing of the eigenfrequency of the
modes of the liquid, which are determined by the denominator of the
"intrinsic" response of (\nnormcomp). This will be so in particular for
small sizes of the slab as determined by our $L$.
For these reasons it is not astonishing that for a thin system
the friction coefficient vanishes like
\equorder{fricpthinf}
\fricpthinf
\edef\nfricpthinf{\the\equnumbera}
Here $\zeta(x)$ is the Riemann
Zeta function, with $\zeta(5)=1.04$.

Conversely, for a {\it large system} the spectrum of the frequencies
may become {\it dense}. Under this condition we expect a {\it finite}
friction coefficient. Indeed, replacing the sum in (\nfricp) by an
integral one finds
\equorder{fricpbin}
\fricpbin
\edef\nfricpbin{\the\equnumbera}
The integration can be carried out to give:
\equorder{fricpbig}
\fricpbig
\edef\nfricpbig{\the\equnumbera}
which is to say {\it we recover the wall formula as the macroscopic limit of
nuclear dissipation}. It should be no surprise that this limit (\nfricpbig)
coincides with the results found in [\nabran, \nabkol] for the a semi-infinite
Fermi liquid.

For later purpose it is worth to remember that here the notion
of a macroscopic limit has been defined in literal sense: It is
obtained by letting the size of the system become larger and larger. In
the next section we will prove that for friction this limit can be
achieved also for a {\it finite} system if we only apply some smoothing
procedure to smear out the discreteness of the "internal" frequencies.

Finally in this section we would like to present some numerical
results.  In Fig.1 we show the friction coefficient calculated from
(\nfricp) as the function of the dimensionless size parameter $\kl /
\pi$. We observe that quite a sizable value of $\kl / \pi$ is necessary
before $\gamfr$ can be said to be represented by the wall formula.

\vskip 10pt
\goodbreak
{\bf \num Frequency smoothing}
\bigskip
\ref{johan}
\edef\njohan{\the\refnumber}
\ref{hofkeszt}
\edef\nhofkeszt{\the\refnumber}
\ref{strut}
\edef\nstrut{\the\refnumber}

We have mentioned already in the previous section, that friction should
not be expected to occur for systems whose modes have a finite spacing. As
suggested in [\njohan] and worked out in more detail later in
[\nhofkeszt] for quantum systems it is necessary to employ some statistical
averaging in the sense of an energy or frequency smoothing before it
becomes physically meaningful to speak of a friction force. The results
of the last section clearly demonstrate that these features hold true
even for our present case, at least for situations with small
$k_\bot L/\pi$. In this paper we do not want to discuss any
justification for such a smoothing. We just want to employ it and
explore the consequences for our present model of a finite slab.

Discussing energy smoothing the Strutinsky method [\nstrut] comes to
one's mind. In [\nhofiva] this connection has been
investigated by studying response functions of finite nuclei, in
particular those which are needed to describe collective (surface)
modes. One should not forget, however, that there is an essential
difference in the two methods. Traditionally the Strutinsky method has
been applied to evaluate the static energy of the system. Applying
essentially the same technique to response functions leads to a drastic
consequence: in this way one introduces {\it irreversibility} into the
dynamics of the system, even in the quantum case. This comes about
whenever the smoothing width
is larger than the level spacing [\nhofkeszt]. This latter condition
necessarily implies that the excitations of the system can no longer be
followed individually but are to be treated statistically.

\ref{strabra}
\edef\nstrabra{\the\refnumber}
In the spirit of [\nhofiva] we define for quantities like
$\gamma({\vec k}_\perp,\omega)$ the frequency average as
\footnote{*}{It should be noted that a similar smoothing procedure has
been applied in [\nstrabra] to derive generalized RPA equations, albeit
without studying the consequences with respect to irreversibility}:
\equorder{enertrfacavd}
\enertrfacavd
\edef\nenertrfacavd{\the\equnumbera}
The $\weightf$ is the weight function,  which
we will choose to be given by a Lorentzian:
\equorder{lorentz}
\lorentz
\edef\nlorentz{\the\equnumbera}
with $\gamma_{\rm av}$ denoting the averaging parameter.
This allows an analytical calculation. From (\nnormcomp) and
(\nenertrfac) one gets:
\equorder{enertrfacav}
\enertrfacav
\edef\nenertrfacav{\the\equnumbera}
The frequency averaged friction coefficient then reads:
\equorder{fricavd}
\fricavd
\edef\nfricavd{\the\equnumbera}

Formula (\nenertrfacav) can be reduced to simpler expressions. We see
two ways to do that and we want to present both. Each one leads to a
different form and both turn out to be useful to discuss the limiting
cases of small or large values of the averaging interval $\gamma_{\rm
av}$. More precisely, as we will see soon, it is the combination
\equorder{avpard}
\avpard
\edef\navpard{\the\equnumbera}
which really matters. Physically it measures the ratio of the averaging
interval to the frequency
\equorder{travfreq}
\travfreq
\edef\ntravfreq{\the\equnumbera}
with which the particles at the Fermi surface traverse the slab back
and forth. The estimate for this quantity given on the right hand side
is obtained by taking for $L$ the diameter of a spherical nucleus
(c.f.[\nsvvm]).

One possibility to study (\nenertrfacav) is offered by the result
(\nenertrfacp) obtained in the previous section if we realize that the
frequency average amounts to calculate the intrinsic response functions
at frequencies having an appropriately chosen, finite and positive
imaginary part. This is to
say we may get the energy averaged friction coefficient by continuing
analytically the form (\nenertrfacp) into the complex plane and by
evaluating it at ${\tilde s}_n =  {\tilde s}_n^\prime + i {\tilde
s}_n^{\prime \prime}$ where ${\tilde s}_n$ is given by (\ndsnwv) with
$\omega=i \gamma_{\rm av}$. One gets:
\equorder{fricavsum}
\fricavsum
\edef\nfricavsum{\the\equnumbera}
We have used the following short hand notation:
\equorder{sized}
\sized
\edef\nsized{\the\equnumbera}
and
\equorder{funcaln}
\funcaln
\edef\nfuncaln{\the\equnumbera}
The expression is also convenient to investigate analytically the
averaged friction coefficient at $\eta \ll 1$.

The other possibility of calculating the friction coefficient from
(\nenertrfacav) is to perform first the integrations over the two
momenta $p_z$ and $p_x$ (or $p_y$ ) and to sum over $n$ afterward.
Details of this calculation will be given in the Appendix. Here we just
state the result:
\equorder{fricavint}
\fricavint
\edef\nfricavint{\the\equnumbera}
This expression allows for an expansion in the small parameter $(d /
\eta )^2 $ to obtain for $\eta \gg 1$:
\equorder{fricavbet}
\fricavbet
\edef\nfricavbet{\the\equnumbera}

It is instructive to study somewhat closer the simpler case of
$k_\perp = 0$. The result can be deduced from
(\nfricavint). It reads:
\equorder{fricavintko}
\fricavintko
\edef\nfricavintko{\the\equnumbera}
Here one can see in analytical fashion that the averaged friction
coefficient is smaller or equal to the wall formula for any value of
the averaging parameter $\eta$
\equorder{fricavleswf}
\fricavleswf
\edef\nfricavleswf{\the\equnumbera}
and that it approaches the latter for $\eta \gg 1$.
For the asymptotic region the deviations can be estimated to be:
\equorder{fricavkobet}
\fricavkobet
\edef\nfricavkobet{\the\equnumbera}

In Fig.2 we present the results of a numerical calculation of the
averaged friction coefficient according to (\nfricavsum) as a
function of the dimensionless averaging parameter $\gamma L / \pi v_F$,
see (\navpard), for two values of $\kl / \pi$. The numerical results
confirm our analytical analysis.

\vskip 10pt
\goodbreak
{\bf \num Conclusions}
\bigskip
We have studied a model for which there exists a macroscopic limit,
if defined in literal sense by letting the size of the system become
larger and larger. We have focussed on the dynamical property of
dissipation. Neglecting effects of collisions entirely, this limit
was found to be given by the wall formula.

As one of our central results, we have been able to prove that this
limit can be obtained for a finite system as well if we only apply an
appropriate frequency averaging. This is interesting for several
reasons. First of all, it is nice to have a model where such a feature
can be shown analytically.  Secondly, it is of particular interest to
be able to do so for a dynamical quantity like the friction
coefficient. Thirdly, and most important are the implications of the
frequency average with respect to real nuclei.  No doubt, there the
situation is different, in the sense that for the small nuclear system
the dynamics of the nucleons must be treated quantum mechanically. Our
present treatment can only be considered to deliver a schematic model
which allows to simulate these effects in analytical fashion.

Having in mind this limitation we may draw some important conclusions,
nevertheless. To begin with, it appears evident that nuclear
dissipation cannot be of pure one-body nature. The presence of the
width $\gamma_{\rm av}$ in formulas like (\nenertrfacav) represents
some uncertainty in the single particle energies of our fluid. In the
formulation  given here this energy width has been introduced by way of
our smoothing procedure. But this $\gamma_{\rm av}$ may also be
understood to simulate effects of couplings from the 1p-1h excitations
to more complicated ones. Indeed, had we taken into account a collision
term in crudest relaxation time approximation the $\gamma_{\rm av}$
would be nothing else but the inverse relaxation time (times $\hbar)$.
Certainly, a realistic treatment of the effects of collisions will be
much more involved. One would at least want to use a relaxation time
ansatz which satisfies the conservation laws. More important may be
improvements which account for the dependence of the effects of
collisions on both frequency as well as temperature.

Let us estimate the typical values needed for $\gamma_{\rm av}$. From
Fig.2 we observe that the $\eta$ defined in (\navpard) should be of
order unity. According to the numbers given in (\ntravfreq) this
implies for $\gamma_{\rm av}$ values of the order of $7 \dots 10 \;
{\rm MeV}$, if converted to energy.  These are values which are not
unreasonably large, in particular if one takes into account that the
energy smoothing should be associated to simulate systems at finite
excitations.

\ref{hoyaje}
\edef\nhoyaje{\the\refnumber}
\ref{yahosa}
\edef\nyahosa{\the\refnumber}
Finally we like to comment briefly on a possible relation of the
macroscopic limit to increasing excitations.  For the case of the
static energy it is a well established fact that this macroscopic limit,
as given by the liquid drop energy, can be reached either by energy
averages or simply by raising the system's temperature $T$. For a
dynamical situation this question is not settled. Features discussed
above show that the problem is more complicated than for the static
situation, last not least because the influence of collisions changes
with excitation energy. First hints we may get from the microscopic
computations of the friction coefficient of finite nuclei published in
[\nhoyaje] and [\nyahosa] (see also [\nhirorev]). There it was found
that for large temperatures the value of friction reaches a constant
value of about half the one of the wall formula. It must be noted that
in these calculations a complicated structure of the effective width
has been considered. This width itself was chosen to saturate at large
excitations.

\vskip 10pt
{\bf Acknowledgements}
One of us (V.I.A.) would like to thank the Deutsche
Forschungsgemeinschaft for financial support and the Physics Department
of the TUM for the kind hospitality extended to him during his stay.
The authors would like to thank F.A.Ivanyuk, A.G.Magner and V.A.Plujko
for stimulating discussions and helpful suggestions.  They gratefully
acknowledge help by D. Kiderlen in preparation of the figures.
%%%%%%%%%%%%%%%%% APPENDIX %%%%%%%%%%%%%%
\vskip 1.5cm
{\bf Appendix}
\bigskip
\newcount\equnumberapp
\equnumberapp=0
\def\eqna{%\global\advance\equnumberapp by 1
         \eqno(A.\the\equnumberapp)}
\def\equorderapp#1{%[[#1
         \global\advance\equnumberapp by 1}
\ref{gradryz}
\edef\ngradryz{\the\refnumber}

In this Appendix we want to present the
derivation of formula (\nfricavint). We begin by noticing that
(\nenertrfacav) can be rewritten as
\equorderapp{ncav}
\ncav
\edef\nncav{\the\equnumbera}
To this expression we apply the analytical continuation discussed in
the text for deriving (\nfricavsum).
Taking into account (\nncav), (\nnormcomp), (\nenertrfac) and
(\nenertrfacp) the averaged friction coefficient (\nfricavd) becomes:
\equorderapp{fricavsumw}
\fricavsumw
\edef\nfricavsumw{\the\equnumberapp}
with the parameters of (\navpard, \nsized, \nfuncaln) and the $w(z)$
defined like in eq.(\nfuncwsn). This form reduces
to the one given in (\nfricavsum) if in the expression for $w(z)$ the
logarithmic representation for the $\rm arccot$ is used.

By means of the integral representation for the function
$w(z)$
\equorderapp{funcwint}
\funcwint
\edef\nfuncwint{\the\equnumberapp}
this expression can be rewritten as:
\equorderapp{fricavsumwi}
\fricavsumwi
\edef\nfricavsumwi{\the\equnumberapp}
if one simply uses the following closed form [\ngradryz]
for the infinite series involved:
\equorderapp{sumclf}
\sumclf
\edef\nsumclf{\the\equnumberapp}
After taking into account the relation
\equorderapp{threl}
\threl
\edef\nthrel{\the\equnumberapp}
and by changing the the integration  variable to $u={\pi \over 2x}$ one
obtains the result (\nfricavint).

%%%%%%%%%%%%%%%%%%%%%%%%APPENDIX%%%%%%%%

\vskip 50pt
\goodbreak
{\bf Figure captions}
\bigskip
Fig.1 The dependence of the friction coefficient (\nfricp)
on the size of the system.

Fig.2 The dependence of the friction coefficient (\nfricavsum) on the
averaging parameter $\eta =\gamma L / \pi v_F$, for two different
values  of the size of the slab, $d = 0.1$ and $d = 1$.

\vfill
\eject

%\vskip 20pt
%\goodbreak
{\bf References}
\bigskip
\item{1)}\lzhirorev
\item{2)}\lzhigoroobn
\item{3)}\lzranswiat
\item{4)}\lzabran
\item{5)}\lzabkol
\item{6)}\lzkoonrand
\item{7)}\lzreif
\item{8)}\lzfeibelman
\item{9)}\lzivan
\item{10)}\lzsije
\item{11)}\lzkidhofiva
\item{12)}\lzbekha
\item{13)}\lzlalifhyd
\item{14)}\lzlalifphyskin
\item{15)}\lzlan
\item{16)}\lzfom
\item{17)}\lziv
\item{18)}\lzholet
\item{19)}\lzhofiva
\item{20)}\lzabkolnp
\item{21)}\lzkolsie
\item{22)}\lzsvvm
\item{23)}\lzjohan
\item{24)}\lzhofkeszt
\item{25)}\lzstrut
\item{26)}\lzstrabra
\item{27)}\lzhoyaje
\item{28)}\lzyahosa
\item{29)}\lzgradryz

\vfill
\end